\documentclass[useAMS,usenatbib,referee]{biom}
\usepackage{multirow}
\usepackage[figuresright]{rotating}
\usepackage{amsmath}
\usepackage{cases}
\usepackage{color}
\usepackage{rotating}
\usepackage{setspace}
\usepackage{graphicx}
\usepackage{subfigure}
\usepackage{float}
\usepackage{epsfig}
\usepackage{lscape}
\usepackage{multirow}
\usepackage{mathrsfs,amssymb}
\usepackage{lscape}
\usepackage{threeparttable}
\usepackage{array,booktabs}
\usepackage{natbib}

\include{GrandMacros}

\DeclareMathAlphabet\mathbfcal{OMS}{cmsy}{b}{n}

\newcolumntype{C}[1]{>{\centering\arraybackslash}p{#1}}
\makeatletter
\newcommand{\specificthanks}[1]{\@fnsymbol{#1}}
\makeatother

\title[Testing small study effects in multivariate meta-analysis]{Testing small study effects in multivariate meta-analysis}

\author{Chuan Hong$^{1*}$,
Georgia Salanti$^{2}$,
Sally Morton$^{3}$,
Richard Riley$^{4}$,
Haitao Chu$^{5}$,
\\ \bf Stephen E. Kimmel$^{6, 7}$,
and Yong Chen$^{7*}$\email{ychen123@pennmedicine.upenn.edu; chong@hsph.harvard.edu}\\\
$^{1}$Department of Biostatistics, Harvard T.H. Chan School of Public Health, Boston, MA, USA \\
$^{2}$Institute of Social and Preventive Medicine, University of Bern, Bern, Switzerland\\
$^{3}$Department of Statistics, University of Virginia Tech, Blacksburg, VA, USA\\
$^{4}$Research Institute for Primary Care \& Health Science, Keele University, Staffordshire, UK\\
$^{5}$Division of Biostatistics, University of Minnesota, Minneapolis, MN, USA\\
$^{6}$Department of Medicine, Perelman School of Medicine, University of Pennsylvania, Philadelphia, PA, USA\\
$^{7}$Department of Biostatistics, Epidemiology \& Informatics, Perelman School of Medicine, \\University of Pennsylvania, Philadelphia, PA, USA
}

\begin{document}

\volume{}
\artmonth{}



\label{firstpage}
\begin{abstract}
Small study effects occur when smaller studies show different, often larger, treatment effects than large ones, which may threaten the validity of systematic reviews and meta-analyses. The most well-known reasons for small study effects include publication bias, outcome reporting bias and clinical heterogeneity. Methods to account for small study effects in univariate meta-analysis have been extensively studied. However, detecting small study effects in a multivariate meta-analysis setting remains an untouched research area.  One of the complications is that different types of selection processes can be involved in the reporting of multivariate outcomes. For example, some studies may be completely unpublished while others may selectively report multiple outcomes. In this paper, we propose a score test as an overall test of small study effects in multivariate meta-analysis. Two detailed case studies are given to demonstrate the advantage of the proposed test over various naive applications of univariate tests in practice. Through simulation studies, the proposed test is found to retain nominal Type I error with considerable power in moderate sample size settings. Finally, we also evaluate the concordance between the proposed test with the naive application of univariate tests by evaluating 44 systematic reviews with multiple outcomes from the Cochrane Database.
\end{abstract}


\maketitle

\section{ Introduction}
In the last several decades, systematic reviews and meta-analyses have received increasing attention in comparative effectiveness research and evidence-based medicine. Meta-analysis is a statistical procedure that combines the results of multiple scientific studies. In meta-analysis, small study effects (SSE) is a well-known critical and challenging issue that may threaten the validity of the results \citep{song2000publication, sutton2000empirical}. ``Small-study effects'' is a generic term for the phenomenon that smaller studies sometimes show different, often larger, treatment effects than large ones \citep{sterne2001funnel}. One of the most well-known reason for SSE is publication bias (PB), in which case the chance of a small study being published does not depend on its quality, but on its effect size, significance or direction.  A possible reason is that authors tend to report significant or positive results or journals tend to publish studies with significant or positive results.  Besides PB, outcome reporting bias (ORB) and clinical heterogeneity (i.e. variability in the participants, interventions and outcomes) in small studies are also important sources for SSE. 

SSE (including PB) is arguably the greatest threat to the validity of meta-analysis \citep{schwarzer2015small}. Erroneous conclusions can arise from a meta-analysis if SSE is not properly accounted for. For example, conclusions from several meta-analyses were later found to be contradicted by mega-trials \citep{egger1995misleading}. In the last two decades, a great deal of effort has been devoted to better reporting protocols, risk of bias evaluation, and statistical methods to detect and correct for SSE based on reported studies \citep{rothstein2006publication, burkner2014testing}. 

Among the many statistical methods on this issue, funnel plots have been commonly used to study SSE. Because the precision of an estimated treatment effect generally increases as the sample size of component studies increases, results from small studies typically scatter widely at the bottom of the funnel plot, while those from larger studies scatter narrowly at the top. An asymmetric inverted funnel is often equated with potential PB \citep{light1984summing, sterne2001funnel}. It is worth mentioning that although a funnel plot is commonly used to detect PB, it should be seen as a generic means of displaying different types of small-study effects, not limited to PB only \citep{egger1997bias, sterne2001funnel}.  However, observations based on funnel plots can themselves be subjective as demonstrated by \citep{lau2006evidence}. Without quantitatively measuring the funnel plot symmetry, different observers may reach different conclusions. Statistical tests based on funnel plot symmetry have been developed, including the rank correlation test \citep{begg1994operating} and regression-based tests \citep{egger1997bias, macaskill2001comparison, harbord2006modified, peters2006comparison, rucker2008arcsine}. Using the symmetry of funnel plots, \cite{duval2000nonparametric, duval2000trim} further developed the nonparametric ``Trim and Fill'' method for imputing missing studies in a meta-analysis. 

Multivariate meta-analysis (MMA), which jointly analyzes multiple and possibly correlated outcomes, has recently received a great deal of attention \citep{jacksonmultivariate}. By borrowing information across outcomes, MMA improves estimation of both pooled effects and between-study variances. However, to the best of our knowledge, statistical tests that quantify the evidence of SSE for multivariate meta-analysis data have not been developed. In fact, there are several unique challenges in MMA that need to be properly addressed in developing a sensible test to study SSE in MMA. 

The first challenge comes from various scenarios of SSE. Unlike univariate meta-analysis, some studies may have only part of their outcomes selectively reported, known as \emph{outcome reporting bias} (ORB) \citep{chan2005identifying}. ORB has received less attention than PB despite its high prevalence in the literature \citep{dwan2008systematic}. More precisely, by comparing trial publications to protocols, \cite{dwan2008systematic} found strong evidence that outcomes that are statistically significant have a higher probability of being reported. More recently, an investigation of the impact of ORB in reviews of rheumatoid arthritis from the Cochrane Database of Systematic Reviews (hereinafter refer to as the ``Cochrane Database'') suggests that ORB has the potential to affect the conclusion in meta-analysis \citep{frosi2015multivariate}. Therefore, a measure to quantify SSE needs to include both PB and ORB scenarios.

The second challenge is to fully account for the multivariate nature of MMA. With multiple outcomes, a common practice is to apply a univariate test, such as the Egger's test, to each of outcomes and report PB for outcomes with small p-values \citep{chan2004empirical, jun2010effects, kavalieratos2016association}.  The problem of multiple testing is often ignored, which may cause excessive false positive findings. For example, \cite{kavalieratos2016association} investigated the association of palliative care with quality of life, symptom burden, survival, and other outcomes for people with life-limiting illness and for their caregivers. In their evaluation of risk of PB, they applied Egger's test to each of the three primary outcomes and yielded p-values of 0.03, 0.09 and 0.37 respectively. And they concluded that quality of life and symptom burden are subject to PB. A more rigorous approach is to apply a Bonferroni correction to the multiple tests. However, as multiple outcomes are often correlated, the tests are correlated and hence the Bonferroni correction is conservative. As lack of power is a well acknowledged concern in detection of PB \citep{turner2013impact}, separately applying a Bonferroni correction to each outcome can further reduce the statistical power \citep{rothstein2006publication}. Thus, in the effort of studying SSE in MMA, we shall aim to apply the strength of MMA by combining information across outcomes.

The third challenge is that within-study outcome correlations, typically required by MMA methods, are often not reported and are difficult to obtain even on request \citep{riley2008alternative, chen2015inference}. As a consequence, the standard likelihood involves unknown within-study correlations, which makes the traditional likelihood based tests (e.g., Wald, score and likelihood ratio tests) not applicable.

In this paper, we propose a score test to study the overall evidence of SSE. To the best of our knowledge, this is the first test for SSE in MMA setting. The proposed test has the following properties.  First, by combining evidence of SSE across multiple outcomes, the proposed test can detect SSE due to PB and/or ORB. Secondly, by jointly modeling multivariate outcomes, the proposed test fully accounts for the multivariate nature of MMA, which avoids the separate investigations of individual outcomes.  We demonstrate the superior power of the proposed test as compared to the simple procedure of separate investigation of outcomes. Furthermore, the proposed test is based on a pseudolikelihood of MMA in the same spirit as \cite{chen2014alternative}. A key advantage is that within-study correlations are not required. Lastly, the test statistic has a closed-form formula and a simple approximated distribution. 

The remainder of this paper is organized as follows. In Section~2, we propose the pseudolikelihood and the corresponding score test. In Section~3, we illustrate the proposed test using two case studies. In Section~4, we conduct simulation studies to compare the proposed test with univariate tests, and investigate the practical implications of the proposed test through an empirical evaluation of 44 systematic reviews from the Cochrane Database. Finally, we provide a brief discussion in Section~5.

\section{Method}\label{sec:model}
In this section, we introduce notation for the multivariate random-effects meta-analysis and review the existing funnel-plot-based methods for detecting {SSE}, including Egger's regression test and two of its variations, Begg's rank test and the Trim and Fill method.

\subsection{Notations for multivariate random-effects meta-analysis}
We consider a meta-analysis with $m$ studies where a common set of $J$ outcomes are of interest. For the $i$th study, let $Y_{ij}$ and $s_{ij}$ denote, respectively, the summary measure for the $j$th outcome and the associated standard error, both assumed known, for $i=1, \ldots, m$ and $j=1,\ldots, J$. Each summary measure $Y_{ij}$ is an estimate of the true effect size $\theta_{ij}$. To account for heterogeneity in effect size across studies, we assume $\theta_{ij}$ to be independently drawn from a common distribution with overall effect size $\beta_j$ and between-study variance $\tau_j^2$, $j=1,\ldots, J$. Under the assumption of a normal distribution for $Y_{ij}$ and $\theta_{ij}$, a multivariate random-effects model is often taken to be \\
\begin{eqnarray}\label{eq:brma}
\left(\begin{array}{c}Y_{i1} \\ \vdots \\Y_{iJ} \end{array} \right)&\sim& N\left( \left(\begin{array}{c}\theta_{i1} \\ \vdots\\ \theta_{iJ} \end{array} \right), {\pmb{\Delta_i}}\right),\quad
{\pmb{\Delta_i}}=\left(\begin{array}{ccc} s_{i1}^2 & \ldots & s_{i1}s_{iJ}\rho_{\textrm{W}_{i{(1J)}}}\\ \vdots & \ddots &\vdots \\s_{iJ}s_{i1}\rho_{\textrm{W}_{i{(1J)}}} & \ldots &s_{iJ}^2 \end{array} \right),
\\
\left(\begin{array}{c}\theta_{i1} \\ \vdots\\ \theta_{iJ} \end{array} \right)&\sim& N\left(\left(\begin{array}{c}\beta_{1} \\ \vdots \\ \beta_{J} \end{array} \right), {\pmb{\Omega}}\right),\quad
{\pmb{\Omega}}=\left(\begin{array}{ccc} \tau_{1}^2 & \ldots & \tau_{1}\tau_{J}\rho_{\textrm{B}{(1J)}}\\ \vdots &\ddots &\vdots \\ \tau_{J}\tau_{1}\rho_{\textrm{B}{(1J)}} & \ldots &\tau_{J}^2 \end{array} \right),
\end{eqnarray}
where ${\pmb{\Delta_i}}$ and ${\pmb{\Omega}}$ are $J\times J$ study-specific within-study and between-study covariance matrices, respectively, and $\rho_{\textrm{W}_{i{(jk)}}}$ and $\rho_{\textrm{B}{(jk)}}$ are the respective within-study and between-study correlations between the $j$th and $k$th outcomes \citep{jacksonmultivariate}. When the within-study correlations $\rho_{\textrm{W}_{i{(jk)}}}$ are known, inference on the overall effect sizes
$(\beta_1, \ldots, \beta_J)$ can be based on the marginal distribution of $(Y_{i1}, \ldots, Y_{iJ})$, i.e.,

\begin{equation}\label{eq:mma_model}
\left(\begin{array}{c}Y_{i1} \\ \vdots \\ Y_{iJ} \end{array} \right)\sim N\left(\left(\begin{array}{c}\beta_{1} \\ \vdots \\ \beta_{J} \end{array} \right), \bf{V_i}\right), \quad
\bf{V_i}={\pmb{\Delta_i}}+{\pmb{\Omega}}.
\end{equation}
We note that the variance of $Y_{ij}$ is partitioned into two parts {($s_{ij}^2$ and $\tau_j^2$)} as in analysis of variance (ANOVA) for univariate random-effects models, and the covariance is also partitioned into two parts as the sum of within- and between-study covariances, i.e., ${\textrm{cov}}(Y_{ij}, Y_{ik})=s_{ij}s_{ik}\rho_{\textrm{W}_{i{(jk)}}}+\tau_{j}\tau_{k}\rho_{\textrm{B}{(jk)}}$. However, study-specific within-study correlations $\rho_{\textrm{W}_{i{(jk)}}}$ among multiple outcomes are generally unknown \citep{riley2008alternative, chen2014alternative}.

\subsection{A score test for multivariate meta-analysis}
Due to the recent development of MMA on multiple outcomes and multiple treatments (e.g., network meta-analysis), the statistical tests based on funnel plot symmetry for univariate outcomes are no longer applicable. In this section, we propose a score test for SSE under MMA. The proposed test can be considered as a multivariate extension of Egger's regression test, which allows for different types of SSE, fully accounts for the multivariate nature of MMA, and does not require within-study correlations. 

A standard score test based on the likelihood function of the model~(\ref{eq:mma_model}) can account for the multivariate nature, as well as borrow information across outcomes. However, the within-study correlations, which are often not reported in the literature, are required in order to calculate the test statistic and to allow for borrowing information across outcomes. Our strategy is to first construct a pseudolikelihood without the use of within-study correlations and then use the corresponding score test for SSE. 

We refer to our test as the \emph{Multivariate Small Study Effect Test} (MSSET) hereinafter. Specifically, we calculate each component of the pseudolikelihood \citep{gong1981pml} using the \emph{estimated} residuals in the regression model of the univariate Egger's test and then employ the idea of composite likelihood \citep{lindsay1988composite} to combine individual pseudolikelihoods together. For the $j$th outcome of the $i$th study, define the standardized effect size as $\textrm{SND}_{ij}=Y_{ij}\left(s_{ij}^2+\tau_j^2\right)^{-1/2}$ and the precision as $\textrm P_{ij}=\left(s_{ij}^2+\tau_j^2\right)^{-1/2}$. We have
\begin{eqnarray} \label{eq:SND_regr}
\textrm{SND}_{ij}&=&a_j+\beta_j\textrm P_{ij}+\varepsilon_{ij},
\end{eqnarray}
where $\varepsilon_{ij}$ is a standard normal random variable. Let $\tilde \tau_j^2$ denote a consistent estimator of $\tau_j^2$ (e.g., a moment estimator). By substituting $\tilde \tau_j^2$ into model~(\ref{eq:SND_regr}), we obtain the log pseudolikelihood
\begin{equation}\label{eq:pseudo_lik2}
\log L_p^j(a_j, b_j)=-{1\over 2}\sum_{i=1}^m {(\widetilde{\textrm{SND}}_{ij}-a_j-b_j \tilde{\textrm{P}}_{ij})^2} ,
\end{equation}
where $\widetilde{\textrm{SND}}_{ij}$ and $\tilde{\textrm{P}}_{ij}$ are simply $\textrm{SND}_{ij}$ and $\textrm P_{ij}$ with $\tau_j^2$ replaced by $\tilde \tau_j^2$. We note here that the regression coefficients $(a_j, b_j)$ (in particular $a_j$) are parameters of interest, while the heterogeneity $\tau_j^2$ is a nuisance parameter. Replacing the nuisance parameter by its estimate can reduce its impact and offer a simple inference procedure. Such an idea was originally proposed by \cite{gong1981pml}, where the uncertainty associated with the estimated nuisance parameters is properly accounted for, and it was later studied by \cite{liang1996abp} and \cite{chen10} in various settings.

To combine the signal for SSE from multivariate outcomes, we propose the following pseudolikelihood by synthesizing individual pseudolikelihoods across outcomes:
\begin{equation}\label{eq:pseudo_lik}
\log L_p({\bf{a}}, {\bf{b}}) =\sum_{j=1}^J \log L_p^j(a_j, b_j),
\end{equation}
where ${\bf{a}}=(a_1, \ldots, a_J)^T$ and ${\bf{b}}=(b_1, \ldots, b_J)^T$.
By simply adding together the log pseudolikelihoods for the various outcomes, we avoid the use of within-study correlations. This idea is similar to the composite likelihood \citep{lindsay1988composite} or independence likelihood \citep{chandler2007inference}, where (weighted) likelihoods are multiplied together whether or not they are independent. An important distinction here is that each component in $\log L_p({\bf{a}}, {\bf{b}})$ is a pseudolikelihood (instead of a likelihood), and its score function is \emph{not} an unbiased estimating equation. Using a similar argument as in the proofs of Lemma~2.1 and Theorem~2.2 in \cite{gong1981pml}, for the $j$th outcome, we have shown the asymptotic consistency and normality of the maximum pseudolikelihood estimator in Appendices~A and ~B of the Supplementary Materials.

With the pseudolikelihood in equation~(\ref{eq:pseudo_lik}), testing SSE in MMA can be carried out by simply testing $H_0: {\bf{a}}=0$, where $a=(a_1, \ldots, a_J)^T$.  We propose the following procedure, where the calculation at each step has a closed-form expression.

\begin{enumerate}
\item {\bf Calculation of the pseudo-score function}
\\
The maximum pseudolikelihood estimator under $a_j=0$ can be calculated as
\begin{eqnarray*}
\tilde b_j(0)&=&\left(\sum_{i=1}^m \tilde{\textrm{P}}_{ij}^2\right)^{-1}\left(\sum_{i=1}^m \widetilde{\textrm{SND}}_{ij} \tilde{\textrm{P}}_{ij}\right).
\end{eqnarray*}
The pseudo-score function w.r.t.\ $\bf a$ under the null can be calculated as ${\bf U}_{\bf a}\left[{\bf 0}, \tilde {\bf b}(\bf 0), \tilde \tau^2\right] = \left(\sum_{i=1}^m \widetilde{\textrm{SND}}_{i1} - \textrm{R}_1, \ldots,  \sum_{i=1}^m \widetilde{\textrm{SND}}_{iJ} - \textrm{R}_J\right)^T$, where $\textrm{R}_j=\tilde b_j(0) \sum_{i=1}^m \tilde{\textrm{P}}_{ij}$. For notation simplicity, we use ${\bf U}_{\bf a}\left[{\bf 0}, \tilde {\bf b}(\bf 0)\right] $ to denote ${\bf U}_{\bf a}\left[{\bf 0}, \tilde {\bf b}(\bf 0), \tilde \tau^2\right]$ hereafter.

\item {\bf Calculation of information matrices}\\
The negative Hessian of the log pseudolikelihood function evaluated at $(\bf{0}, \tilde {\bf b}(\bf 0))$ is calculated as
    \begin{eqnarray*}
    {\bf I}_0=
   \left(
   \begin{array}{cc}
   {\bf I_{0_{aa}}}& {\bf I_{0_{ab}}}\\
   {\bf I_{0_{ab}}}^T    &{\bf I_{0_{bb}}}
    \end{array}
    \right),
    \end{eqnarray*}
    where
${\bf I_{0_{aa}}}=\textrm{diag}(m, \ldots, m)$ is a $J$-dimensional diagonal matrix  with $m$ as its diagonal elements,
${\bf I_{0_{ab}}}=\textrm{diag}(\sum_{i=1}^m P_{i1}, \ldots, \sum_{i=1}^m P_{iJ})$ is a $J$-dimensional diagonal matrix with $\sum_{i=1}^m P_{ij}$ as its $j$th element, and ${\bf I_{0_{bb}}}=(\sum_{i=1}^m P_{i1}^2, \ldots, \sum_{i=1}^m P_{iJ}^2)$ is a $J$-dimensional diagonal matrix with $\sum_{i=1}^m P_{ij}^2$ as its $j$th element. The $J\times J$ submatrix of the inverse of ${\bf I}_{0}$ with respect to $\bf a$, denoted by ${\bf I}_0^{aa}$, can be calculated as $\left(\bf I_{0_{aa}}-{\bf I}_{0_{aa}} {\bf I}_{0_{bb}}^{-1} {\bf I}_{0_{ab}}\right)^{-1}$.

\item {\bf Calculation of the test statistic} \\
Let ${\pmb \Sigma}_{aa}$ denote the asymptotic variance of $\sqrt{m} (\hat {\bf a}-{\bf a}_0)$.
The calculation of ${\pmb \Sigma}_{aa}$ requires properly accounting for the additional uncertainty in $\tilde {\pmb \tau}^2$, as we described in Appendix C of the Supplementary Materials. Let $\bar \lambda$ denote the  arithmetic mean of the eigenvalues of $({\bf I}_{0}^{aa})^{-1}\bf \Sigma_{aa}$.  The proposed MSSET test is constructed by
\begin{equation}\label{msset}{\textrm{MSSET}}={\left(m \bar \lambda\right)}^{-1} {\bf U}_{a}\left[{\bf 0}, \tilde {\bf b}({\bf 0})\right]^T{{\bf I}_0}^{aa}{\bf U}_{a}\left[{\bf 0}, \tilde {\bf b}({\bf 0})\right].\end{equation}
The test statistic is compared with the $\chi_J^2$ distribution to obtain a $p$-value.
\end{enumerate}
A proof of the asymptotic distribution of the score test is provided in Appendix~C of the Supplementary Materials.  We note that instead of constructing the test following the traditional score test for composite likelihood (e.g., the pseudo-score statistics defined on page 193 in \cite{molenberghs2005models}), the above test is constructed for better computational stability. Finally, the proposed test reduces to the traditional Egger's test for univariate meta-analysis when the number of outcomes is one (i.e., $J=1$). The above steps for obtaining the MSSET and $p$-value will be illustrated in Section~4.1.

\subsection{A modified version of MSSET for multivariate meta-analysis with binary outcomes.} 

The proposed MSSET test, along with the Egger's regression, assume that, under the null hypothesis of no SSE, there is no association between effect size and precision. However, this does not hold for binary outcomes. For example, if a binary outcome is summarized by the log-odds ratio (logOR), the variance estimators of logOR are statistically dependent of the estimated logOR. This may induce inflated type I errors.  To reduce the correlation between the effect size and precision, PetersÕ test, HarbordÕs score test, and RuckerÕs Arcsine-Thompson (AS-Thompson) test have been proposed for randomized clinical trials with balanced sample sizes. \cite{jin2014modified} proposed a regression method using a smoothed variance as the precision scale of an individual study to test for publication bias, and suggested that the smoothed regression method is more robust across different settings. In the same spirit as in \cite{jin2014modified}, we modified the MSSET test by replacing the within-study variance $s_{ij}$ in equation (3) by its smoothed version. 

For illustration, consider a bivariate meta-analysis of $n$ individual studies, where the $j$th outcome is binary summarizing in logOR. For the $i$th study, let $a_i$ and $b_i$ denote numbers of cases and controls in the exposed group, and $c_i$ and $d_i$ denote numbers of cases and controls in the unexposed group. We obtain the logOR and its variance as $\log (a_i d_i/b_ic_i)$, and $1/a_i+1/b_i+1/c_i+1/d_i$, which are intrinsically correlated. To reduce the correlation, the estimated smoothed variance for the estimated logOR is given by 
$$\left\{(a_i+c_i )p_1/n\right\}^{-1}+\left\{(a_i+c_i )(1-p_1)/n\right\}^{-1}+\left\{(b_i+c=d_i )p_0/n\right\}^{-1}+\left\{(b_i+d_i )(1-p_0)/n\right\}^{-1},$$
where $p_1=\sum_{i=1}^n \left\{a_i/(a_i+c_i)\right\}$, and $p_0=\sum_{i=1}^n \left\{b_i/(b_i+d_i)\right\}$.

\section{Two case studies}
In this section, we consider two case studies: 1) structured telephone support or non-invasive tele-monitoring for patients with heart failure and 2) prognostic value of MYCN and Chromosome 1p in patients with neuroblastoma. 

\subsection{Structured telephone support or non-invasive tele-monitoring for patients with heart failure}
Heart failure is a complex, debilitating disease. To improve clinical outcomes, and reduce healthcare utilization, specialized disease management programs are conducted, such as structured telephone support and non-invasive home telemonitoring. Over the last decade, trials have been conducted to examine the effects of these programs. To compare structured telephone support or non-invasive home tele-monitoring interventions with standard practice, \cite{inglis2016structured} conducted a systematic review including 41 randomized clinical trials. Primary outcomes included all-cause mortality, and both all-cause and heart failure-related hospitalizations. Other outcomes included length of stay, health-related quality of life, heart failure knowledge and self-care, acceptability and cost. 

We revisit the systematic review conducted by \cite{inglis2016structured}, and use the primary outcome (all-cause mortality) and the secondary outcome (mental quality of life) to illustrate the proposed method. We analyze the 35 trials that reported either of the two outcomes.  Among the 35 trials, 11 reported both outcomes; 34 reported the primary; and 12 reported the secondary outcome.

\begin{figure}[!h]
\centering\includegraphics[scale=0.5]{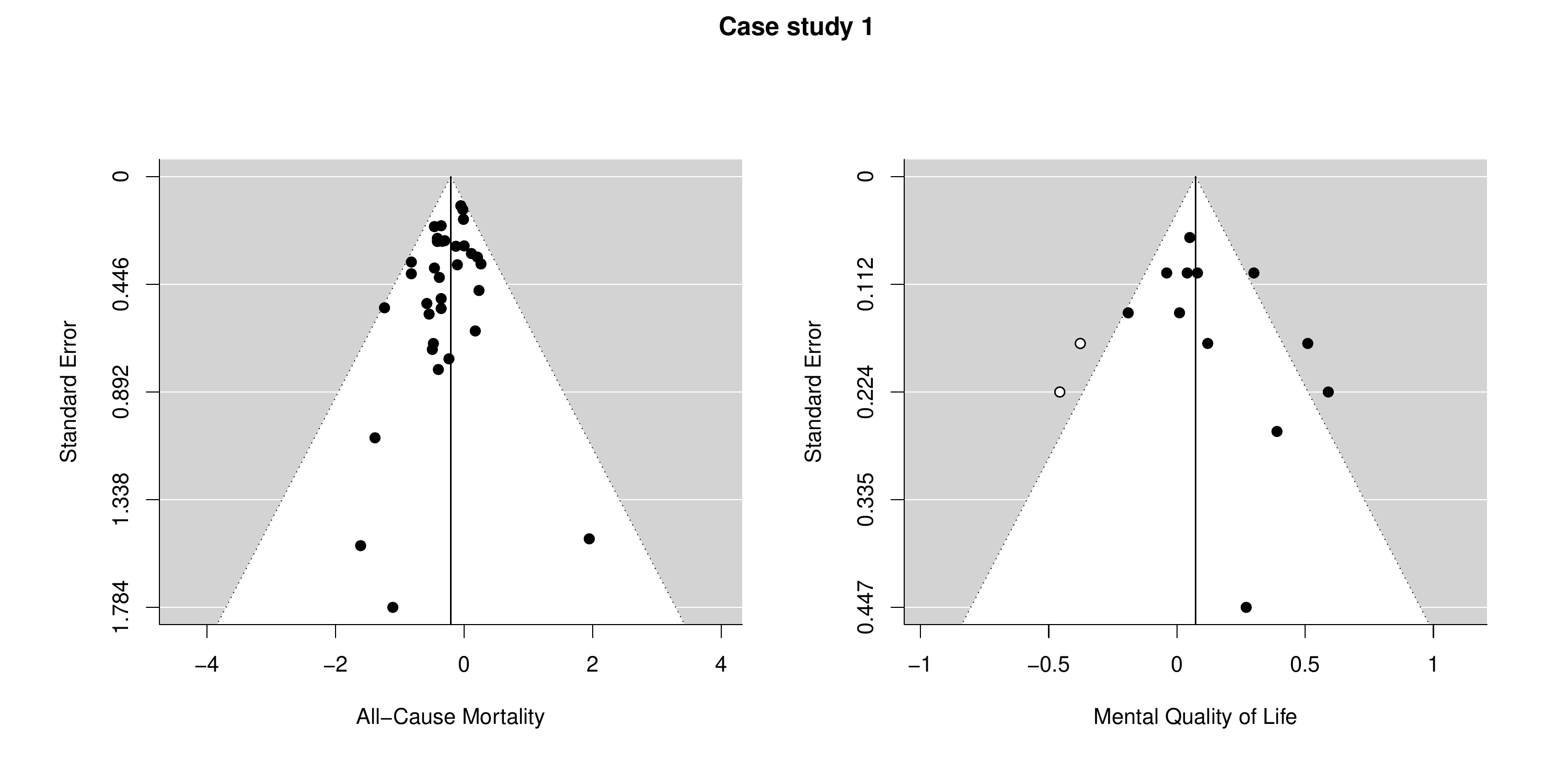}
\centering\includegraphics[scale=0.5]{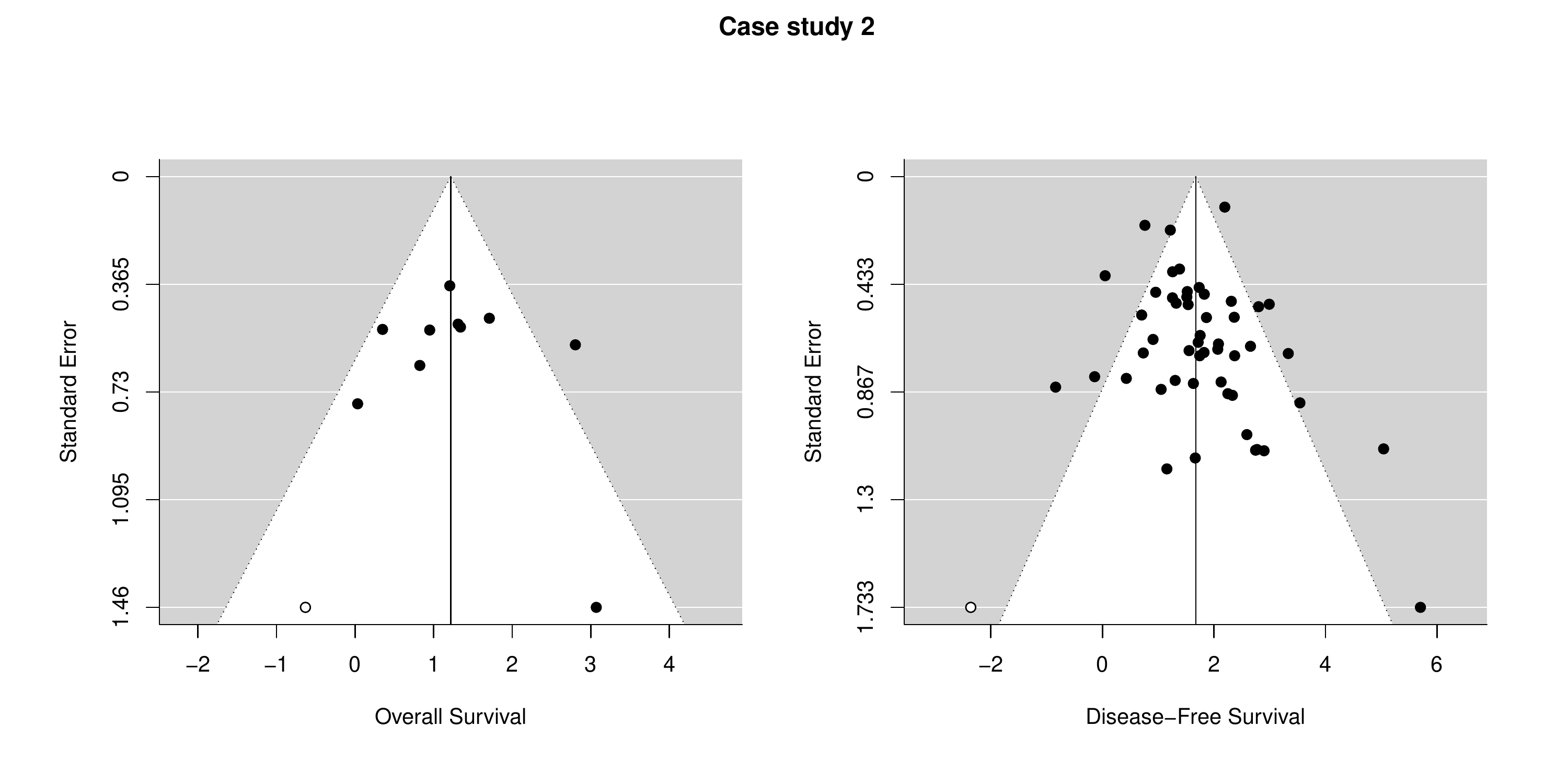}
\caption{Funnel plots for Case study 1 in Section 4.1 (upper panels) and Case study 2 in Section 4.2 (lower panels). }
\label{fig:data}
\end{figure}

{We now illustrate the steps in the proposed MSSET. First we obtain ${\bf b}({\bf 0})=(-0.20, 0.13)^T$. The pseudo-score function under the null hypothesis is ${\bf U}_{\bf a}\left[{\bf 0}, \tilde {\bf b}{\bf{(0)}}\right]=(-4.84, 1.32)^T$. We then calculate the negative Hessian of the log pseudolikelihood,
\begin{eqnarray*}
{\bf I}_0=\left(
\begin{array}{cccc}
41.00 &  91.46& 0.00 &  0.00\\
91.46&  324.40 & 0.00   &0.00\\
 0.00   &0.00 &41.00& 57.54\\
0.00 &  0.00 &57.54 &292.79
\end{array}
\right).
\end{eqnarray*}
The $2\times 2$ submatrix of the inverse of ${\bf I}_0$ w.r.t.\ $\bf a$ is calculated as
\begin{eqnarray*}
{\bf I}_0^{aa}=\left(
\begin{array}{cc}
6.57 \textrm{E}-02&0\\
0&3.37\textrm{E}-02
\end{array}
\right).
\end{eqnarray*}
We then get \
\begin{eqnarray*}
{\bf \Sigma}_0^{aa}=\left(
\begin{array}{cc}
7.10\textrm{E}-04&-2.79\textrm{E}-07\\
-2.79\textrm{E}-07&1.10\textrm{E}-05
\end{array}
\right),
\end{eqnarray*}
and $\bar \lambda=0.05$. The proposed test statistic MSSET can be obtained from equation~(\ref{msset}) as $7.02$. By comparing with $\chi_2^2$, the $p$-value is found to be $0.03$.

The upper panel in Figure~\ref{fig:data} displays the funnel plots of all-cause mortality and mental quality of life of 35 trials. The funnel plots of both all-cause mortality and quality of life suggest severe asymmetry.  Such an observation is confirmed by univariate tests of SSE (Egger's test), as we summarize in the upper panel in Table~\ref{tab:case studies}. When we apply EggerÕs test to the two outcomes separately, both are marginally significant (p=0.09 and 0.06). But after Bonferroni correction, both are not significant.  {In contrast, the proposed test yields statistically significant result with smaller p-values. We use this example to highlight a limitation of univariate tests: the loss of power after adjusting for multiple testing. In contrast, the proposed MSSET test is better in identifying SSE by combining signals of SSE from multiple outcomes.

\begin{table}[!h]
\centering
\caption{$p$-Values of testing SSE using Egger's test, Begg's test, and the proposed score test (msset).} 
\begin{tabular}{cccc}
\hline
           &       Test & Outcome 1 & Outcome 2 \\
\hline
Case study 1 &      Egger &      0.09 &     0.06  \\

           &       MSSET &      \multicolumn{2}{c}{0.03 }          \\
           \\
           \\
Case study 2 &      Egger &   0.01 &      0.58\\
           &       MSSET &      \multicolumn{2}{c}{0.02}     \\
\hline
\end{tabular}
\label{tab:case studies}
\end{table}

\subsection{MCYN and Chromosone 1p in patients with neuroblastoma}
{Neuroblastoma, a type of cancer that starts in early nerve cells, is a common extracranial solid tumor of childhood \citep{huang2013neuroblastoma}. One of the best-characterized genetic marker of risk in neuroblastoma is amplification of MYCN, which is usually used as a prognostic indicator.  Studies have been conducted to assess whether amplified levels of MYCN and deletion of chromosome 1p, Ch1p, are associated with survival outcomes in children with neuroblastoma. \cite{jackson2011multivariate} conducted a meta-analysis to examine the association of the two factors (MYCN and Chromosome 1p) with overall and disease-free survival. Seventy-three studies assessing the prognostic values of MYCN and Chromosome 1p, in patients with neuroblastoma are included in this meta-analysis. Up to four estimates of effect are provided by the individual studies, including an estimated unadjusted log hazard ratio of survival, either of the high relative to the low level group of MYCN, or Chromosome 1p deletion to its presence. 

We study the SSE for overall survival and disease free survival using Egger's test and the proposed MSSET test. The lower panel in Figure~\ref{fig:data} displays the funnel plots of the two outcomes. For the overall survival, we observe a certain degree of asymmetry, while for the disease-free survival, we do not observe such evidence. Similarly, as shown in the lower panel in Table~\ref{tab:case studies}, applying univariate methods for SSE to outcomes separately only lead to the detection of SSE for overall survival. The proposed MSSET test suggests a statistically significant evidence of overall SSE for bivariate outcomes. 

\section{Simulation studies and an empirical evaluation using 44 systematic reviews}
In this section, we evaluate the performance of the proposed MSSET test through fully controlled simulation studies, and study the empirical impact of our new MSSET.
\subsection{Simulation studies}
{In our simulation studies, we compare the proposed MSSET test with the commonly used Egger's regression test and Begg's rank test.} The data are generated from MMA {with a common set of two outcomes} as specified by model (\ref{eq:brma}). To cover a wide spectrum of scenarios, we vary the values {of} several factors that are considered important in practice: 1) To reflect the heterogeneity in the standard error of the summary measure across studies, we sample $s_{ij}^2$ from the square of the distribution $N(0.3,0.5)$, which leads to a mean value of $0.33$ for $s_{ij}^2$. 2) The size of the within-study variation relative to the between-study variation may have a substantial impact on the performance of the methods. To this end, we let the between-study variances $\tau_1^2=\tau_2^2$ range from 0.1 to 2 to represent the random-effects model with relatively small to large random effects. 3) For within-study and between-study correlations, we consider $\rho_{\textrm{W}i}$ to be constant with value $-0.5$, $0$, or $0.5$ and $\rho_{\textrm{B}}$ to be constant with value $-0.5$, $0$, or $0.5$. 4) The number of studies $n$ is set to 10, 25, 50, or 100 to represent meta-analysis of small to large numbers of studies.
We consider the Type I error setting where the selection (i.e., the decision of whether to publish a particular study or report a particular outcome) does not depend on the effect sizes. We also consider the power settings where the selection depends on the effect sizes.  We conduct {5000} simulations for the Type I error setting and {1000} simulations for the power settings. For the power settings, in order to obtain $n$ published studies, we follow three steps: 1) $N$ studies are simulated, where $N$ is an integer greater than $n$ (here we choose $N=3n$ to ensure there are enough studies before selection); 2) studies are excluded based on four different selection scenarios described in the following paragraph; and 3) $n$ studies are sampled randomly from those that remain after the previous step.
\subsubsection{Selection model scenarios}
To evaluate the usefulness and performance of the proposed MSSET test in detecting the SSE in multivariate meta-analysis setting, it is critical to design and cover a wide range of publication/reporting scenarios, which we refer to as selection model scenarios. We consider four different scenarios for two types of SSE, including scenarios where studies are completely missing (denoted by Scenario C1, C2, and C3), and scenarios where studies are partially missing (denoted by Scenario P).

For Scenarios C1--C3, the probability of a study being published depends on the $p$-values of both effect sizes of this study. For example, as shown in the upper left panel of Figure~\ref{fig:scenarios}, a study is published if and only if the $p$-values of both effect sizes are less than 0.05. Although Scenarios C2 and C3 may be more common in practice, we include Scenario C1 as it was usually considered in the literature as a benchmark for sanity check of a procedure \citep{burkner2014testing}. The scenarios C2 and C3 in the upper right and lower left of Figure~\ref{fig:scenarios}, can be thought as a ``smoothed'' version of scenario C1 where the publication of a study relies on, but is not totally decided by the significance of two outcomes. 

For Scenario P (P stands for partial reporting), as visualized in the lower-right panel of Figure~\ref{fig:scenarios}, the probability of an outcome being reported depends on the $p$-value of its effect size; therefore, some studies may selectively report only one of the outcomes. In addition, the selection model in Scenario P is different from those in Scenarios C1--C3 in that we let the logit of the probability of the $j$th outcome in the $i$th study being published be $(-2.5+0.1\textrm{SND}_{ij}+1.5\textrm{SND}_{ij}^2)I(\textrm{SND}_{ij}<2)+4I(\textrm{SND}_{ij}\ge2)$. This selection model is empirically estimated from a meta-analysis where the true status of a registered study being published or not is available \citep{turner2008selective}, making this scenario most practically plausible. A brief description, as well as the data of obtaining this selection model are provided in Appendix D of the Supplementary Materials.

\begin{figure}[!t]
\centering\includegraphics[scale=0.6]{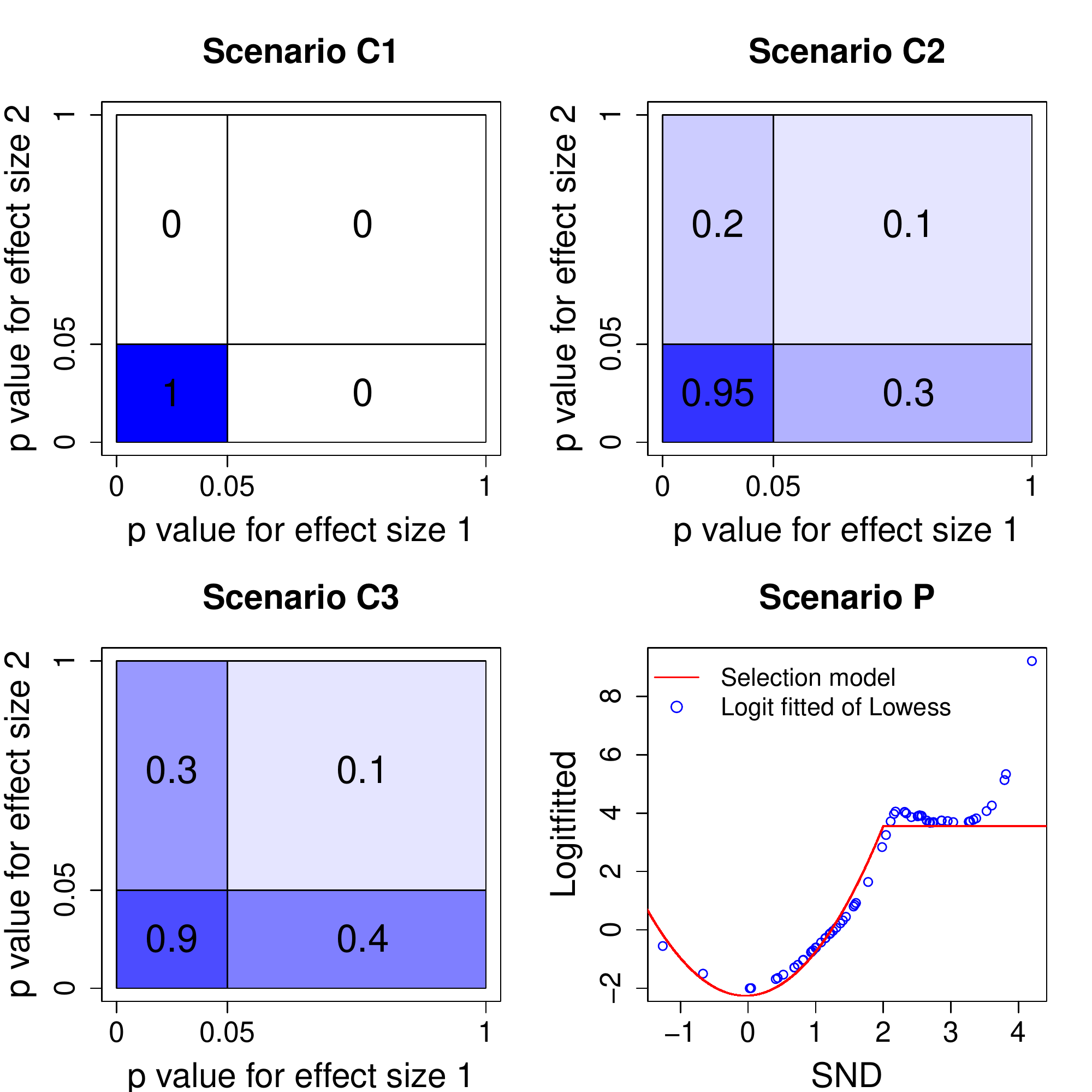}
\caption{Probability of being published under Scenarios C1, C2, C3, and P. In Scenarios C1-C3, the probabilities are shaded from dark to light (i.e., the largest probability refers to the darkest shade).} 
\label{fig:scenarios}
\end{figure}

\subsubsection{Simulation results}
Table~\ref{tab:sim} summarizes the Type I errors at the $10\%$ nominal level of the tests under comparison. The proposed MSSET test controls the Type I errors well in all settings. {The Egger's regression test for one outcome only (Egger$_1$) or two outcomes separately with Bonferroni correction (Egger) have slightly inflated Type I errors when the sample size is small ($n=10$). } We observe that the Type I errors of Begg's rank test are very conservative when the sample size is relatively small but inflated when it is relatively large. This observation is consistent with the literature, where investigators report that Begg's rank test does not perform well in controlling Type I errors \citep{burkner2014testing}.

\begin{table}[!t]
\centering
\caption{Type I errors ($\times 100\%$), at the $10\%$ nominal level, of Egger's test, Begg's test, and the proposed MSSET test. The univariate tests are conducted on the first outcome only, denoted as ``$test_1$'', and the Bonferroni correction is used to combine the test results, denoted as ``test''.  The number of studies $n$ is 10, 25, 50, 75, and 100, and the between-study heterogeneity $\tau^2$ is 0.1, 0.9, and 1.9.} 
\begin{tabular}{ccccccc}
\hline
n   & $\tau^2$ & MSSET & Egger$_1$ & Egger & Begg$_1$ & Begg \\
\hline
10  & 0.1    & 12.9 & 14.7   & 15.8  & 2.4   & 2.1  \\
    & 0.9    & 12.7 & 13.2   & 14.6  & 2.6   & 2.0  \\
    & 1.9    & 11.9 & 13.1   & 15.0  & 2.3   & 1.6  \\
25  & 0.1    & 11.3 & 11.9   & 13.0  & 17.8  & 22.4 \\
    & 0.9    & 11.1 & 11.4   & 12.2  & 24.1  & 32.7 \\
    & 1.9    & 10.9 & 11.3   & 12.0  & 24.6  & 33.7 \\
50  & 0.1    & 10.4 & 10.7   & 10.9  & 38.1  & 51.8 \\
    & 0.9    & 10.5 & 10.8   & 10.6  & 44.7  & 61.7 \\
    & 1.9    & 10.4 & 10.9   & 10.6  & 45.5  & 62.8 \\
75  & 0.1    & 10.0 & 9.8    & 9.8   & 48.2  & 65.5 \\
    & 0.9    & 10.3 & 10.2   & 9.8   & 55.9  & 73.8 \\
    & 1.9    & 10.1 & 10.5   & 10.2  & 56.6  & 74.8 \\
100 & 0.1    & 9.9  & 9.9    & 9.6   & 55.4  & 72.1 \\
    & 0.9    & 9.7  & 9.8    & 9.7   & 62.0  & 79.8 \\
    & 1.9    & 9.7  & 10.2   & 9.6   & 63.0  & 80.7\\
\hline
\end{tabular}
\label{tab:sim}
\end{table}

\begin{figure}[!t]
\centering\includegraphics[scale=0.6]{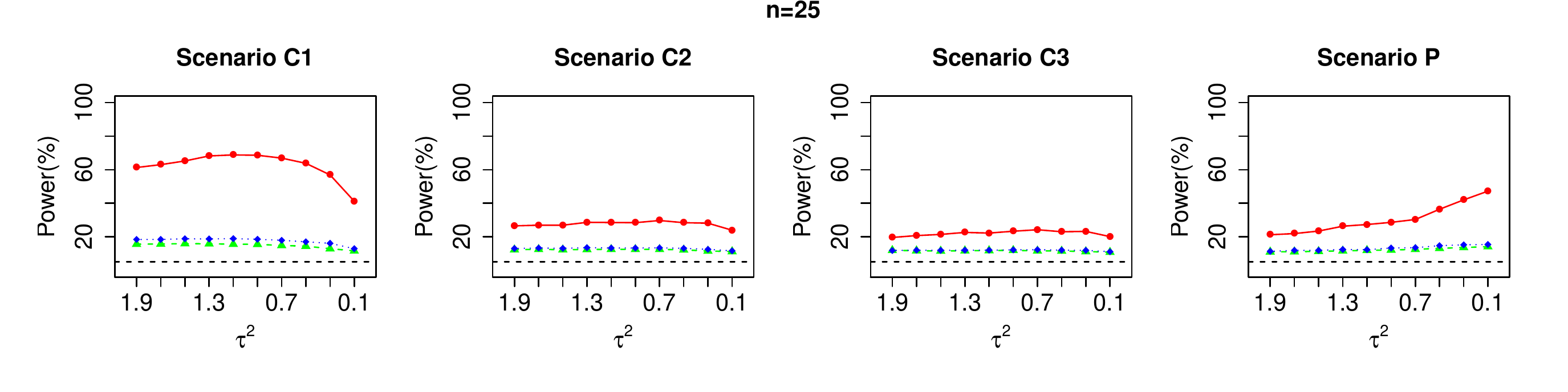}
\centering\includegraphics[scale=0.6]{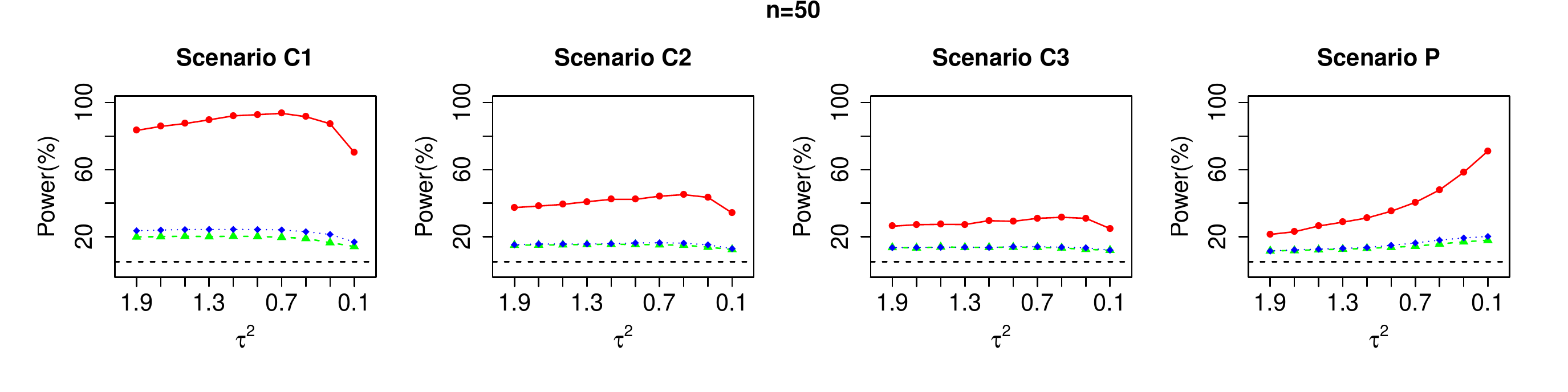}
\centering\includegraphics[scale=0.6]{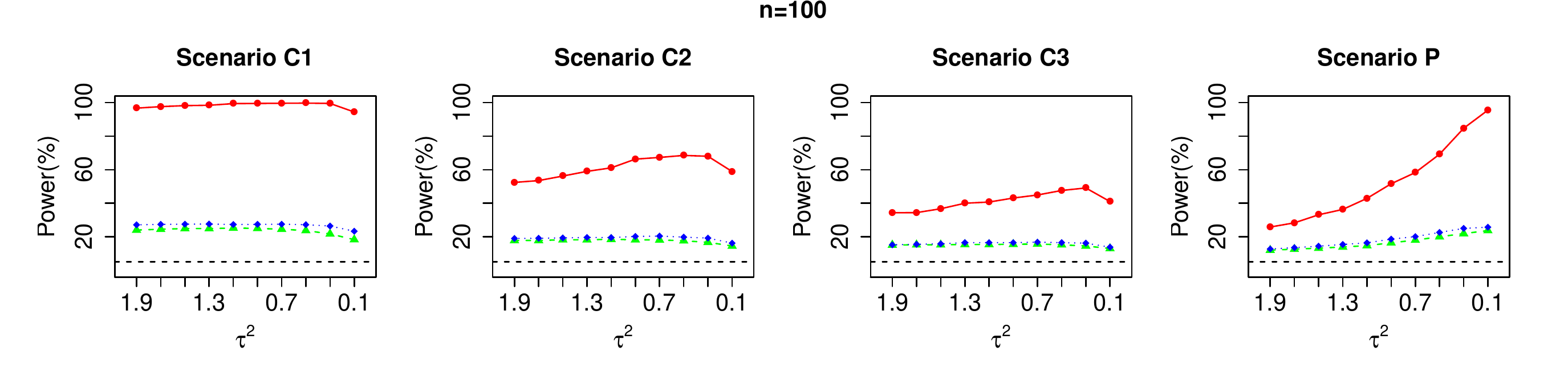}
\centering\includegraphics[scale=0.6]{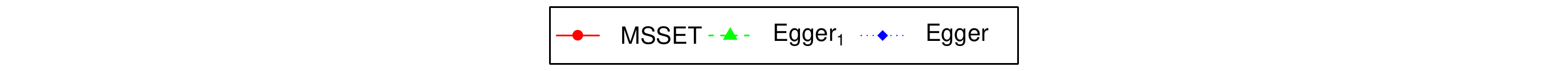}

\caption{{Power plots of the proposed MSSET test, Egger's regression test on one outcome (Egger$_1$), and Egger's regression test on two outcomes with Bonferroni correction (Egger) at the $10\%$ nominal level for sample sizes varying from 25 to 100 and between-study variances varying from 0.25 to 5.}} 
\label{fig:power}
\end{figure}
Figure~\ref{fig:power} summarizes the power of the tests under comparison. {Clearly, the proposed MSSET test is the most powerful under all the settings considered. Egger's regression test on one outcome (Egger$_1$) and that on two outcomes with Bonferroni correction (Egger) have substantial power loss compared with the proposed MSSET.} There are several additional interesting findings from Figure~\ref{fig:power}:

1) For Scenarios C1--C3, we observe non-monotonic trends for the MSSET test and Egger's regression test. A possible explanation for this non-monotonicity is due to the fact that both the MSSET test and Egger's regression test assume the random-effects model and require the estimation of the between-study variance $\tau^2$. However, when the between-study variance is close to zero, a fixed-effect model is more suitable than a random-effects model. Therefore, these two tests have lower power as heterogeneity is very small, as a consequence of lack of good fit assumed for the random-effect meta-analysis model. 

2) Unlike in Scenarios C1--C3, the power curve in Scenario P is increasing. One possible explanation is that this scenario allows selective reporting of parts of multiple outcomes. Because the proposed test combines signals of SSE across multiple outcomes, the number of studies required to identify the same degree of heterogeneity is smaller than that required for the other scenarios.

3) We observe that there is a decreasing trend in power from Scenario C1 to Scenario C3 for all tests under comparison. This indicates that it is easier to detect SSE when selectivity is greater.

\subsubsection{Additional results for multivariate meta-analysis with binary outcomes}
To evaluate the performance of the modified version of MSSET using smoothed within-study variance for binary outcomes as described in Section 2.3 (referred as MSSET$_{\textrm{smooth}}$ hereafter), we consider additional settings where a bivariate meta-analysis has one continuous outcome and one binary outcome. We compare the MSSET$_{\textrm{smooth}}$ test with different versions of Egger's regression method, including the Egger's test on the first outcome only using smoothed variance (denoted as Egger$_{1,\textrm{smooth}}$),  Bonferroni correction of Egger's test on both outcome using smoothed variance (denoted as Egger$_{\textrm{smooth}}$) , and the Egger's test using original variance. We only consider the Type I error setting, because the correlated effect size and original within-study variance are suggested to lead to inflated Type I errors. Table~\ref{tab:sim2} summarizes the Type I errors at the $10\%$ nominal level of the tests under comparison. In general, the proposed MSSET test controls the Type I errors better than the various versions of Egger's tests under different sample size settings.  The Egger's tests with original variance produce inflated Type I errors. This observation is consistent with the literature, where the correlated effect size and its variance will lead to false positive testing results of SSE.  The Egger's tests using smoothed variance control the Type errors better than those using naive variances. 

\begin{table}[!t]
\centering
\caption{Type I errors ($\times 100\%$), at the $10\%$ nominal level, of the proposed MSSET test using smoothed variance (MSSET$_{\textrm{smooth}}$), the Egger's test on the first outcome only using smoothed variance (Egger$_{1,\textrm{smooth}}$), and using the original variance (Egger$_1$), the Egger's test on both outcomes using Bonferroni correction and using smoothed variance (Egger$_{\textrm{smooth}}$), and using the original variance (Egger). The number of studies $n$ is 10, 25, 50, 75, and 100, and the between-study heterogeneity $\tau^2$ is 2.5.}
\begin{tabular}{cccccc}
\hline
n   & MSSET$_{\textrm{smooth}}$ & Egger$_{1, {\textrm{smooth}}}$ & Egger$_1$ & Egger$_{\textrm{smooth}}$& Egger \\
\hline
10  & 13.7    & 13.6      & 13.9   & 14.9     & 15.0  \\
25  & 11.2    & 12.4      & 12.6   & 12.9     & 12.9  \\
50  & 10.2    & 11.1      & 11.8   & 11.3     & 11.6  \\
75  & 10.6    & 10.0      & 11.8   & 10.3     & 11.0  \\
100 & 10.3    & 8.7       & 10.5   & 10.0     & 11.7 \\
\hline
\end{tabular}
\label{tab:sim2}
\end{table}

\subsection{An empirical evaluation using 44 systematic reviews from the Cochrane Database}
To evaluate the practical impact of the proposed MSSET, we compare the results of the proposed multivariate test of SSE with the results from univariate tests of SSE by applying the proposed MSSET test and the Egger's test to a large number of comparable meta-analyses obtained from the Cochrane Database of Systematic Reviews, an online collection of regularly updated systematic reviews and meta-analyses of medical studies. We do not include the Begg's test for comparison, since our simulation study in the previous session indicated that Begg's rank test does not perform well in controlling Type I errors. We extracted {5320} out of {11401} available reviews from the database before July 2015 ({6081} files were corrupt). Among these records, we identified {1675} meta-analyses that compare treatment to placebo or no treatment. The following criteria were then applied: a) the number of studies in a meta-analysis must be at least 10; b) all studies in a meta-analysis must contain a common set of two different outcomes; and c) in any meta-analysis, at least one study should report both outcomes. Similar criteria have been used in the literature \citep{trikalinos2014empirical}. After imposing the above quality control, we obtained 44 meta-analyses with bivariate outcomes.

We compare the results from the proposed MSSET test with the results of the univariate Egger's test on bivariate outcomes separately. Specifically, for each meta-analysis, we apply the proposed MSSET test, the Egger's test on outcome 1 (Egger$_1$), and the Egger's test on outcome 2 (Egger$_2$). We then dichotomize the $p$-value at the level of 0.10, for MSSET test,  Egger$_1$ and Egger$_2$. In addition, we use Bonferroni correction to combine the results from Egger's test of bivariate outcomes (Egger$_{\textrm{Bonferroni}}$). We cross-tabulate the dichotomized results, comparing the MSSET test results with one of the univariate tests. Table~\ref{tab:44} shows 2$\times$2 tables of the number of meta-analyses that identified SSE via the MSSET test, versus one of the three Egger's tests. 

{Among the 44 meta-analyses considered in the analysis, the proposed MSSET test has identified 7 ($16\%$) meta-analyses with SSE at the significance level of 0.10.  Egger's regression test has identified 4 ($9\%$) meta-analyses having SSE\@ for both outcome 1 and outcome 2. For both outcomes with Bonferroni correction, Egger's regression test and Begg's rank test has identified 6 ($14\%$) meta-analyses having SSE\@. The percentage of SSE identified by the MSSET test is the highest among all tests, which is consistent with the simulation results showing that the MSSET test is the most powerful among all tests under comparison.} In addition, the proposed MSSET is in a larger concordance with the Egger$_{\textrm{Bonferroni}}$ test, by combining signals from both outcomes. 

In summary, this ``meta-meta'' analysis of 44 reviews from the Cochrane Database demonstrates the practical implications of the proposed MSSET by comparing the test to the results of univariate tests. The MSSET has consistently detected more SSE than the univariate tests.

\begin{table}[!h]
\centering
\caption{Contingency table of the proposed MSSET test vs.\ Egger's regression based on outcome 1 (Egger$_1$) only and outcome 2 only (Egger$_2$), and Egger test using Bonferroni correction to combine the results from Egger's test of bivariate outcomes (Egger$_{\textrm{Bonferroni}}$) (i.e., S=1 if $p$-value $<0.1$; S=0 if $p$-value $>0.1$).} \begin{tabular}{cccccccccc}
\hline
           & \multicolumn{ 2}{c}{Egger$_1$} &            & \multicolumn{ 2}{c}{Egger$_2$} &            & \multicolumn{ 2}{c}{Egger$_{\textrm{Bonferroni}}$}       \\
\cline{2-3} \cline{5-6} \cline{8-9} 
MSSET & S=0  & S=1 &  & S=0  & S=1 &  & S=0  & S=1  \\
S=0    & 31 & 3 &  & 31 & 3 &  & 31 & 3  \\
S=1    & 6  & 1 &  & 6  & 1 &  & 4  & 3\\
\hline
\end{tabular}
\label{tab:44}
\end{table}

\section{Discussion}
MMA is becoming more common and has the potential to make an important contribution to evidence-based medicine \citep{jacksonmultivariate}. Studying and quantifying the overall evidence of SSE in MMA is of critical importance for evaluating the validity of systematic reviews \citep{rothstein2006publication, burkner2014testing}. In this paper, we have proposed a rigorous score test for studying overall evidence of SSE in MMA. To the best of our knowledge, the proposed test is the first effort to study SSE in an MMA setting. It can be thought of a multivariate extension of Egger's regression test, which naturally is Egger's test when the number of outcomes is one (i.e., $J=1$). For more general settings ($J\geq 2$), the proposed test has the following advantageous properties, besides its simplicity. First, by combining signals of SSE across multiple outcomes (whether or not they are of different data types or on different scales), the proposed test can quantify SSE under different scenarios, and is consistently more powerful than univariate tests.  Second, by jointly modeling multiple outcomes, the score test can fully account for the multivariate nature of MMA, which avoids the need of a Bonferroni correction for multiple testing. As a technique and practical advantage, the within-study correlations are not required in the proposed test, while the between-study correlations are properly accounted for in the testing procedure, by the theory of composite likelihood. 

{Outcome reporting bias (ORB) had not received sufficient attention until recently \citep{chan2005identifying, dwan2013systematic}. \cite{dwan2013systematic} reviewed the evidence from empirical cohort studies assessing ORB and showed that statistically significant outcomes are more likely to be selectively reported. \cite{copas2014model}  suggested a likelihood-based model that reflects the empirical findings to estimate the severity of ORB.  More recently, {\cite{frosi2015multivariate} suggested that the difference between the results from MMA and those from univariate meta-analysis indicates the presence of ORB\@}. Designing a specific test for detecting ORB is a topic of future research.}

A limitation of the proposed method is that it only detects for SSE and does not correct for it. An important extension is to develop a robust procedure to correct for SSE by jointly analyzing multivariate outcomes. Another interesting extension of the proposed test is to network meta-analysis, where multiple treatments are compared jointly in clinical trials but each trial may compare only a subset of all treatments.

To summarize, we have developed a simple and useful test to detect SSE in a multivariate meta-analysis setting. As a natural extension of the univariate Egger's regression test, our test has the advantage of combining signals across outcomes without requiring within-study correlations. We have found that the proposed test is substantially more powerful than the univariate tests and has a practical impact on real applications, calling for more attention on potential SSE or PB during research synthesis. We believe this test is a useful addition for tackling the problem of SSE and PB in comparative effectiveness research.

\backmatter

\bibliographystyle{biom} \bibliography{mPB_MMA}

\label{lastpage}

\end{document}